\documentclass[aps,a4paper,amsmath,amssymb,twocolumn,
floatfix,superscriptaddress]{revtex4}
\usepackage{bm,amsmath,amssymb,amsfonts}
\usepackage[dvips]{graphicx}
\usepackage{color}
\definecolor{dblue}{rgb}{0.0,0.0,0.7}
\definecolor{dred}{rgb}{0.9,0.0,0.0}
\definecolor{dgreen}{rgb}{0,0.6,0.0}
\usepackage{hyperref}

\hypersetup{
    pdfnewwindow=true,      
    colorlinks=true,        
    linkcolor=blue,         
    citecolor=dgreen,       
    filecolor=blue,         
    urlcolor=blue           
}
\begin{document}

\title{Flat bands and nontrivial topological properties in an extended Lieb lattice}
\author{Ankita Bhattacharya}
\email[E-mail: ]{ankita.bhattacharya@tu-dresden.de}
\affiliation{Institute of Theoretical Physics, Technische Universit{\"a}t Dresden, 
01062 Dresden, Germany}
\author{Biplab Pal}
\thanks{Corresponding author}
\email[E-mail: ]{ biplab@post.bgu.ac.il}
\altaffiliation[Present address: ]{Raymond and Beverly Sackler School of Physics and Astronomy, Tel Aviv University, Tel Aviv 69978, Israel.}
\affiliation{Department of Physics, Ben-Gurion University, Beer Sheva 84105, Israel}
\begin{abstract}
We report the appearance of multiple numbers of completely flat band states in an extended Lieb lattice 
model in two dimensions with five atomic sites per unit cell. We also show that this 
edge-centered square lattice can host intriguing topologically nontrivial phases when intrinsic 
spin-orbit (ISO) coupling is introduced in the microscopic description of the corresponding tight-binding 
Hamiltonian of the system. This ISO coupling strength acts like a complex next-nearest-neighbor hopping 
term for this model and can be, in principle, tuned in a real-life experimental setup. In the presence of this 
ISO coupling, the band spectrum of the system gets gapped out, leading to nonzero integer values of the spin 
Chern number for different bands, indicating the nontrivial topological properties of the system. Furthermore, 
we show that for certain values of the ISO coupling, nearly flat bands with nonzero Chern numbers emerge in 
this lattice model. This opens up the possibility of realizing interesting fractional quantum spin Hall 
physics in this model when interaction is taken into account. This study might be very useful in an 
analogous optical lattice experimental setup. A possible application of our results can  also be anticipated 
in the field of photonics using single-mode photonic waveguide networks.  
\end{abstract}
\maketitle
\section{Introduction}
In recent years, the study of novel topological phases of matter has emerged as one of the most exciting 
areas of research in the field of condensed-matter physics. It ties the fundamental physics and technology, 
providing a possible test bed for exploring new kinds exotic particles, such as anyons~\cite{stern-aop08}, 
Majorana fermions~\cite{kane-prl08, alicea-rpp12}, fractons~\cite{nandkishore-arcmp19}, etc., for potential technological 
applications like high-performance electronics and topological quantum computation~\cite{nayek-prl05}. 
Motivated by the theoretical formulation~\cite{kane-mele-prl05, bernevig-prl06}, and subsequent experimental 
observation of the quantum spin Hall effect in HgTe/CdTe quantum wells~\cite{molenkamp-sci07}, intensive research 
has been performed in this direction. In particular, topological insulators have become a hot topic in 
the condensed-matter community owing to their unusual physical properties as well as for their promising 
technological applications in spintronics and quantum computation~\cite{wang-bookchapter16, wang-spin16}.
Topological insulators (TIs)~\cite{moore-nature10, hasan-kane-rmp10, zhang-rmp11} are a special kind of insulator, 
which possesses an insulating band gap in the bulk like a conventional insulator but supports gapless 
conducting states at its edges or surfaces. These unique edge states of TIs are topologically protected by 
time-reversal symmetry and robust against any nonmagnetic and geometric perturbations as long as the bulk gap 
is not closed. Such exotic new states of quantum matter in both two and three dimensions are characterized by 
a topological invariant $\mathbb{Z}_{2}$, which is known as the topological quantum number~\cite{kane-prl07}. 
Often, simple tight-binding models allow us to bring out the essence of interesting topological properties in 
condensed-matter systems. To date, a wide range of tight-binding lattice 
models~\cite{franz-prb09, kaisun-prl09, fiete-prb10} have been probed successfully to evince the existence of 
interesting topological insulating phases of matter in simple lattice geometries.  

On the other hand, translationally invariant lattice systems which exhibit one or more flat bands (FBs) in their Bloch 
spectrum have generated considerable interest over the course of time~\cite{mielke-jpa91, das-sharma-prl11, tang-prl11, 
titus-prl11, flach-prb13, flach-epl14, flach-prl14, flach-prb15, flach-prl16, vicencio-pra17, flach-prb17, ajith-prb18, ajith-prb17, 
biplab-prb18, pal-prb18, jun-won-prb19, nandy-pla19}. The presence of these momentum-independent zero-dispersion bands in the spectrum 
implies the existence of a macroscopic number of entirely degenerate single-particle states at the flat-band energy. Due to 
the vanishing group velocity corresponding to the FBs, wave transport in the system is completely suppressed, leading to a 
strong localization of the eigenstates. In fact, these localized states span only a few lattice sites, forming a 
compact localized state (CLS)~\cite{ajith-prb17, biplab-prb18, pal-prb18, jun-won-prb19, nandy-pla19}. The study of such 
FB systems has always kept scientists intrigued as it provides an ideal playground to investigate various 
interesting strongly correlated phenomena, such as unconventional Anderson localization~\cite{goda-prl06, shukla-prb10}, 
Hall ferromagnetism~\cite{tasaki-prl92, richter-prl12}, high-temperature superconductivity~\cite{heikkila-prb16}, and 
superfluidity~\cite{torma-natcom15, torma-prl16}, to name a few. In recent times, certain interesting 
studies have shown the transition between the flat bands and the Dirac points appearing in simple tight-binding lattice models, 
such as the two-band checkerboard model and three-band kagome and Lieb models~\cite{montambaux-prl18, jiang-prb19, montambaux-arxiv19}. 
Also, systems exhibiting nearly FBs with nonzero Chern 
numbers are proposed to be promising candidates to realize fractional Chern insulators analogous to the flat degenerate 
Landau levels appearing in a continuum~\cite{das-sharma-prl11, tang-prl11, titus-prl11, pal-prb18, franz-prb12, liu-prl12}. 

Interest in FB systems is not restricted only to the condensed-matter community; it has also gained substantial 
attention in the optics domain with recent experimental advances. Due to the high degeneracy and complete localization, 
FBs have relevance in many technological applications, such as diffraction-free propagation of 
light~\cite{vicencio-prl15, mukherjee-prl15}, enhanced light-matter interaction, generating slow 
light~\cite{baba-nat-photon08}, etc. Although the theoretical proposal on the existence of FBs in certain periodic lattice 
geometries has been known for a long time, interest has been renewed in recent times with the experimental realization of 
FBs in a variety of photonic lattices~\cite{vicencio-prl15, mukherjee-prl15, mukherjee-ol15, longhi-ol14, xia-ol16, 
zong-oe16, mukherjee-prl18, ajith-prl18}, ultracold atoms in optical lattices~\cite{jo-prl12, takahashi-sciadv15, 
takahashi-prl17}, and exciton-polariton condensates~\cite{masumoto-njp12, baboux-prl16}. Also, from the experimental point 
of view, in particular, for optics experiments, nonlinear effects are relatively easy to probe in a system with nondispersive 
bands. Apart from these, recently people have come up with new flat-band models based on a real organic material 
platform~\cite{chen-jmca18, ni-nanoscale18, su-apl18, jiang-nanoscale19, zhang-prb19, jiang-natcomm19}, 
for which they have discussed the appearance of intriguing flat-band states adopting density functional theory 
based calculations as well as tight-binding analysis.
\begin{figure}[ht]
\includegraphics[clip, width=0.8\columnwidth]{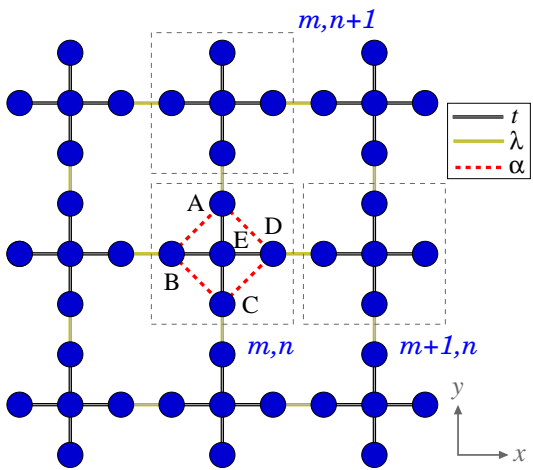}
\caption{Schematic diagram of an extended Lieb 
lattice model representing an edge-centered square lattice with 
five atomic sites per unit cell. The 
indices $(m,n)$ indicate the position of $(m,n)$-th unit 
cell of the lattice structure in the $x$-$y$ plane. Different 
lines represent different kinds of hopping parameters, viz., 
$t$ (intra-cell hopping amplitude), $\lambda$ (inter-cell 
hopping amplitude), and $\alpha$ (next-nearest-neighbor 
complex hopping parameter or the ISO coupling strength), 
respectively, as shown in the inset.}
\label{fig:Lieb-lattice}
\end{figure}

Over the course of time, a wide variety of lattice models have been reported to exhibit FBs in their band structure, among 
which the Lieb lattice has been one of the most popular, and it has been explored with vigor to accomplish various interesting 
physical properties of this lattice model~\cite{torma-prl16, franz-prb12, vicencio-prl15, mukherjee-prl15, xia-ol16, ajith-prl18, 
takahashi-sciadv15, takahashi-prl17}. A conventional Lieb lattice has three atomic sites per unit cell and can be thought to 
originate out of a square lattice. So a natural question which crosses our mind is that, Can we cook up 
other interesting variations of such lattice geometries out of a square lattice with intriguing topological properties? 
In this paper, we address this question and study the topological properties of an edge-centered square lattice with five 
atomic sites per unit cell, named an \textit{extended} Lieb lattice ~\cite{jiang-natcomm19, zhang-aop17} (see 
Fig.~\ref{fig:Lieb-lattice}). We analyze the band spectrum of this lattice geometry based on a tight-binding approach and 
show that it can support multiple FB states in the presence of only nearest-neighbor hopping. Upon inclusion of intrinsic 
spin-orbit (ISO) coupling in the model, it shows interesting nontrivial topological properties. We demonstrate that this model 
can have a nonzero topological invariant, and may act as a possible host for a two-dimensional topological insulator.

In what follows, we present the model and describe the essential results of this study. The rest of the paper is organized as 
follows: In Sec.~\ref{model}, we depict the lattice model and write down the corresponding tight-binding Hamiltonian describing the 
model. In Sec.~\ref{generating-FB}, we illustrate the FBs appearing for this model and also furnish the necessary system parameters 
for their appearance. The role of the inclusion of ISO coupling in the model is explained in detail in Sec.~\ref{effect-iso}. 
The appearance of the corresponding nontrivial topological properties in the system is described in Sec.~\ref{topo-properties} with 
a demonstration of the Berry curvature for the bands with nonzero Chern numbers. In addition, we also show the appearance of the 
edge states in the system confirming its topological properties. In Sec.~\ref{expt-setup}, we discuss the possible implementation of the 
lattice model in an analogous optical lattice setup which provides a very flexible environment for observing complex condensed-matter 
phenomena. Finally, in Sec.~\ref{conclu}, we conclude with a summary of the key findings and future outlook.   
\section{The model and the mathematical scheme}
\label{model}
We start by writing down the tight-binding Hamiltonian for the extended Lieb lattice structure shown in 
Fig.~\ref{fig:Lieb-lattice}. The building block of the lattice structure is repeated periodically over a two-dimensional plane to 
form the entire lattice geometry. The unit cell consists of five atomic sites. The Hamiltonian of the system within a 
tight-binding formalism reads
\begin{align}
\bm{\hat{H}}_{0} = &\sum_{m,n} \Big[ \sum_{i} \tilde{c}^{\dagger}_{m,n,i}\tilde{\epsilon}_{i}\tilde{c}_{m,n,i} \nonumber\\
&+ \sum_{\langle i,j \rangle}\Big(\tilde{c}^{\dagger}_{m,n,i}\tilde{\Lambda}_{ij}\tilde{c}_{m,n,j} + \textrm{H.c.} \Big)\Big],
\label{eq:hamil0-wannier}
\end{align}
where $(m,n)$ stands for the unit cell index and $i$ denotes the atomic site index within a unit cell. 
$\tilde{c}^{\dagger}_{m,n,i} = \big(c^{\dagger}_{m,n,i,\uparrow} \ c^{\dagger}_{m,n,i,\downarrow} \big)$ is the creation 
operator matrix for electrons (with spin {\it up} ($\uparrow$) and spin {\it down} ($\downarrow$), respectively) at site $i$ within cell $(m,n)$. 
$\tilde{\epsilon}_{i}=\textrm{diag}\big(\epsilon_{i,\uparrow}, \epsilon_{i,\downarrow}\big)$ refers to the on-site energy matrix, 
$\tilde{\Lambda}_{ij}=\textrm{diag}(t, t)$ is the intra-cell hopping amplitude matrix, and 
$\tilde{\Lambda}_{ij}=\textrm{diag}(\lambda, \lambda)$ is the inter-cell hopping amplitude matrix. By applying the Fourier transformation 
to the real-space Hamiltonian~\eqref{eq:hamil0-wannier}, we obtain the Hamiltonian in the momentum space $\mathbf{k}\equiv(k_{x},k_{y})$ as
\begin{equation}
\bm{\hat{H}}_{0}=\sum_{\mathbf{k}} \mathbf{\hat{\Psi}_{k}^{\dagger}} \left[\mathbb{\hat{I}}_{2\times2} \otimes 
\bm{\mathcal{\hat{H}}}_0(\mathbf{k})\right] \mathbf{\hat{\Psi}_{k}},
\label{eq:hamil-mom} 
\end{equation}
where $\mathbf{\hat{\Psi}_{k}} \equiv (\mathbf{\hat{\Psi}}_{\mathbf{k},\uparrow} \ \mathbf{\hat{\Psi}}_{\mathbf{k},\downarrow})^{T}$ with 
$\mathbf{\hat{\Psi}}_{\mathbf{k},s} = (c_{\mathbf{k},A,s} \ c_{\mathbf{k},B,s} \ c_{\mathbf{k},C,s} \ c_{\mathbf{k},D,s} \ c_{\mathbf{k},E,s})$. 
It should be noted that, since the two spin projections (spin {\it up} ($\uparrow$) and spin {\it down} ($\downarrow$)) are 
time-reversal partners, 
i.e., $\bm{\mathcal{\hat{H}}}_{0}^{\downarrow}(\mathbf{k}) = [\bm{\mathcal{\hat{H}}}_{0}^{\uparrow}(-\mathbf{k})]^{\ast}$, 
it is sufficient to restrict our attention to any one of the spin projections. We set 
$\bm{\mathcal{\hat{H}}}_{0}^{\downarrow}(\mathbf{k}) = [\bm{\mathcal{\hat{H}}}_{0}^{\uparrow}(-\mathbf{k})]^{\ast} = 
\bm{\mathcal{\hat{H}}}_{0}(\mathbf{k})$. The entire Hamiltonian consists of two uncoupled blocks corresponding to spin-up and 
spin-down projections, and the resulting bands in the $\mathbf{k}$ space are two-fold degenerate. 
The matrix $\bm{\mathcal{\hat{H}}}_{0}(\mathbf{k})$ is given by
\begin{equation}
\bm{\mathcal{\hat{H}}}_{0}(\mathbf{k})=
\left(\def\arraystretch{1.5} \begin{matrix}
\epsilon_{A} & 0 & \lambda e^{ik_{y}} & 0 & t  \\
0 & \epsilon_{B} & 0 & \lambda e^{-ik_{x}} & t \\
\lambda e^{-ik_{y}} & 0 & \epsilon_{C} & 0 & t \\
0 & \lambda e^{ik_{x}} & 0 & \epsilon_{D} & t  \\
t & t & t & t & \epsilon_{E} \\
\end{matrix}\right).
\label{eq:hamil-k}
\end{equation}
We choose $\epsilon_{i,\uparrow} = \epsilon_{i,\downarrow} = \epsilon_{i} = 0$ $\forall$ $i$; $i \in \{A,B,C,D,E\}$. Note that, 
as we incorporate the ISO coupling term in the Hamiltonian later on, we adopt a general spin dependence in our notations 
from the beginning. We extract the important information about the band structure of the system by diagonalizing the matrix in 
Eq.~\eqref{eq:hamil-k}. Here, we have the two tuning parameters $\lambda$ and $t$ in $\bm{\mathcal{H}}_{0}(\mathbf{k})$, which we 
can adjust to have interesting band features for this model. The results are presented in the next section.
\begin{figure}[ht]
\includegraphics[clip, width=0.8\columnwidth]{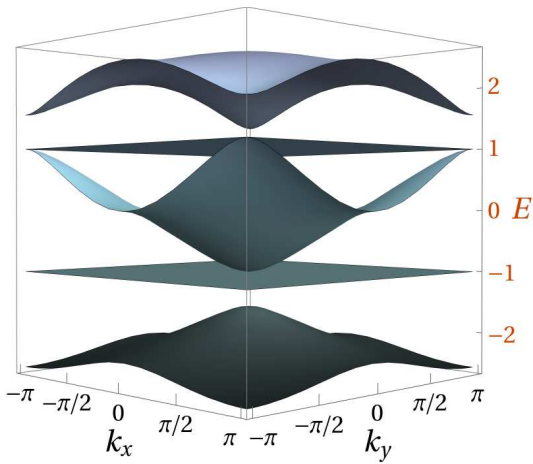}
\caption{Band dispersions for the extended Lieb lattice geometry 
shown in Fig.~\ref{fig:Lieb-lattice}. Two nondispersive completely 
flat bands appear in the band structure at energies $E=\pm 1$. 
We have set the values of the hopping parameters as $t=1$ and 
$\lambda=1$.}
\label{fig:FB-without-SO}
\end{figure}
\section{Formation of the nondispersive flat bands}
\label{generating-FB}
In this section, we show and explain the appearance of FBs in our proposed lattice model. The starting point is to diagonalize 
Eq.~\eqref{eq:hamil-k}. The resulting band spectrum is shown in Fig.~\ref{fig:FB-without-SO}, which comprises two completely flat 
bands and three dispersive bands. The geometrical structure of a lattice is one of the primary reasons for the existence of FBs, 
and is related to the wave localization due to destructive interference on the lattice. This lattice geometry offers such a scenario 
and we can clearly observe the appearance of multiple FBs in its band structure (see Fig.~\ref{fig:FB-without-SO}). Here, we have two 
FBs appearing symmetrically at the energies $E_{\textrm{FB}}=\pm \lambda$, and they are touched by a dispersive band sandwiched in 
between them. This is in contrast to the band structure of a conventional Lieb lattice, where there is one FB 
sandwiched in between two dispersive bands, forming a triply degenerate Dirac point which resembles a spin-$1$ conical-type energy 
spectrum~\cite{jiang-prb19, franz-prb10}. We note that, the band structure and the flat bands appearing here in the case of the 
extended Lieb model cannot be reduced to the band spectrum of a conventional Lieb model via an adiabatic continuous deformation, as 
mentioned in Refs.~\cite{jiang-prb19, montambaux-arxiv19}.

\begin{figure}[ht]
\includegraphics[clip, width=0.8\columnwidth]{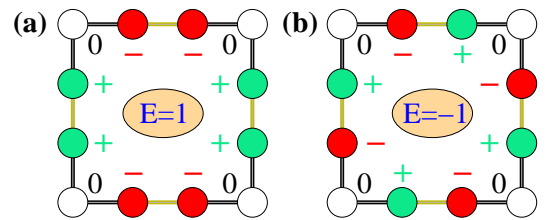}
\caption{Plots of wave-function amplitude distribution corresponding 
to compact localized states (CLS) with FB energies (a) $E=1$ and 
(b) $E=-1$. The green sites indicate positive ($+$) weight, 
red sites indicate negative ($-$) weight, and white sites represent zero ($0$) 
weight. The values of the hopping amplitudes are $t=1$ and $\lambda=1$, 
respectively.}
\label{fig:amp-distribution}
\end{figure}

The appearance of these FBs can be physically interpreted from different points of view. In a momentum space description, we can 
say that the effective mass of the particle in a flat band becomes infinite, making it superheavy such that it cannot move and 
therefore the resulting band is nondispersive. On the other hand, from a real-space point of view, we can say that the hopping 
of the particle between different parts of the lattice is effectively turned off, leading to the formation of 
CLSs~\cite{ajith-prb17, biplab-prb18, pal-prb18}. In the CLS, the particles are highly localized over a few lattice sites, 
and beyond that region, the wave-function amplitude vanishes, as depicted in Fig.~\ref{fig:amp-distribution} for this lattice 
structure.  
The particles in such a setting are self-localized over an octagonal-ring block induced by destructive interference. We may call 
such blocks {\it localized prisons} in which the particles are trapped, circulating over an octagonal-ring zone and not able 
to escape from it. The particles in such localized prisons with no kinetic energy form the flat-band state. We can always find 
a construction of wave-function amplitudes with positive, negative, and zero weights corresponding to a FB state in absence of 
any external magnetic field, as verified for a wide range of lattice geometries elsewhere~\cite{flach-epl14, ajith-prb17, biplab-prb18, 
pal-prb18, jun-won-prb19}. This can be considered an important hallmark of a FB state in any tight-binding lattice model, as is 
the case for the present lattice geometry as well. We note that, one may construct similar CLSs corresponding 
to a general Lieb $(2N+1)$ lattice model supporting $N$ numbers of FBs, where $N$ is the number of sites on the 
edges~\cite{jiang-natcomm19, zhang-aop17}.

The two nondispersive FB states for this lattice model appear at the energies $E_{\textrm{FB}}=\pm \lambda$. This indicates 
that, one can easily control the energy at which the particle freezes by suitably tuning the value of the hopping 
parameter $\lambda$. The gap between the top and bottom dispersive bands and the adjacent flat bands can be controlled by 
changing the value of $t$. The above facts can have a significant implication when one talks about the photonic flat-band 
localization. In a photonic version of the present lattice geometry, by suitably changing the corresponding control parameters, 
one might easily manipulate the frequency at which light will be trapped. This might be useful for technological applications in 
photonics. Up to this point we have discussed the FB states for the present lattice model with only nearest-neighbor 
hopping. Next, we address the effect of adding a complex next-nearest-neighbor hopping to this model. This is encountered by 
including an ISO interaction term in the corresponding Hamiltonian of the system, as discussed in the next section.
 \section{Effect of the intrinsic spin-orbit coupling on the band structure}
\label{effect-iso}
The aim of this section is to see what happens to the band structure of the present lattice geometry when 
an ISO coupling term is included in the Hamiltonian in Eq.~\eqref{eq:hamil0-wannier}. Such an ISO coupling term 
plays the role of an effective magnetic field for the two different spin projections of a particle in a tight-binding 
lattice model. However, there is an important fundamental difference between the two quantities: a real magnetic 
field is an external quantity which breaks the time-reversal symmetry (TRS) of the system, whereas the ISO coupling 
is an intrinsic property of the system and it preserves the TRS. An important physical consequence of this fact is 
that, one can realize the quantum spin Hall effect in a material in the absence of an external magnetic field, as predicted 
by Kane and Mele for a hexagonal lattice model~\cite{kane-mele-prl05}, and subsequently taken forward by others for a 
variety of interesting tight-binding lattice geometries~\cite{franz-prb10, morais-smith-prb12}. Another important motivation 
for the investigation of such a term is that, inclusion of this term often lead to engrossing topological phase transitions in 
the system. To determine if this lattice model supports such interesting topological phases, we include the ISO 
interaction term in the Hamiltonian~\eqref{eq:hamil0-wannier} with the following form: 
\begin{equation}
\bm{\hat{H}_{\textrm{ISO}}} = i\alpha \sum_{m,n} \sum_{\langle\langle i,j \rangle\rangle}
\Big(\nu_{ij}\tilde{c}^{\dagger}_{m,n,i}\sigma_{z}\tilde{c}_{m,n,j} + \textrm{H.c.} \Big),
\label{eq:hamil-iso}
\end{equation}
where $\nu_{ij} = - \nu_{ji} = \pm 1$ according to the orientation of the path connecting the two next-nearest-neighbor sites 
$i$ and $j$~\cite{kane-mele-prl05}. $\alpha$ determines the strength of the ISO coupling. $\sigma_{z}=\textrm{diag}(1, -1)$ 
is the $z$-component of the Pauli spin matrices, and thus the sign of the hopping amplitude is opposite for the two spin 
components, $\uparrow$ and $\downarrow$. 

In the momentum space the corresponding Hamiltonian reads
\begin{equation}
\bm{\mathcal{\hat{H}}_{\textrm{ISO}}} = \pm
\left(\def\arraystretch{1.2} \begin{matrix}
0 & i\alpha & 0 & -i\alpha & 0 \\
-i\alpha & 0 & i\alpha & 0 & 0 \\
0 & -i\alpha & 0 & i\alpha & 0 \\
i\alpha & 0 & -i\alpha & 0 & 0 \\
0 & 0 & 0 & 0 & 0 \\
\end{matrix}\right).
\label{eq:hamil-iso-mom}
\end{equation}
Here, the $+(-)$ sign refers to the Hamiltonian block for the spin-up (spin-down) projection. This effectively means that 
the two different spin species are subjected to two opposite effective magnetic fluxes.  
\begin{figure}[ht]
\includegraphics[clip, width=0.49\columnwidth]{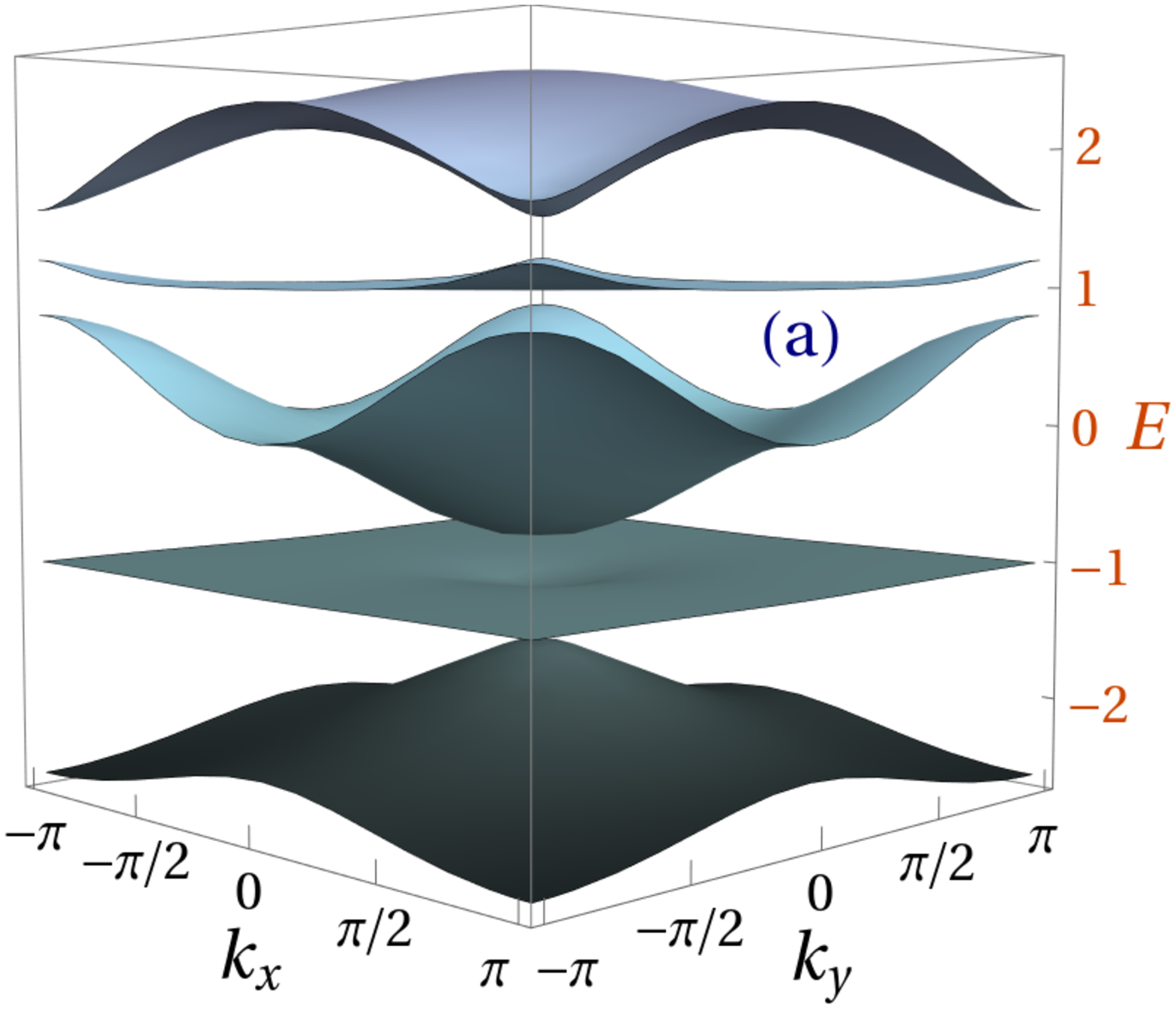}
\includegraphics[clip, width=0.49\columnwidth]{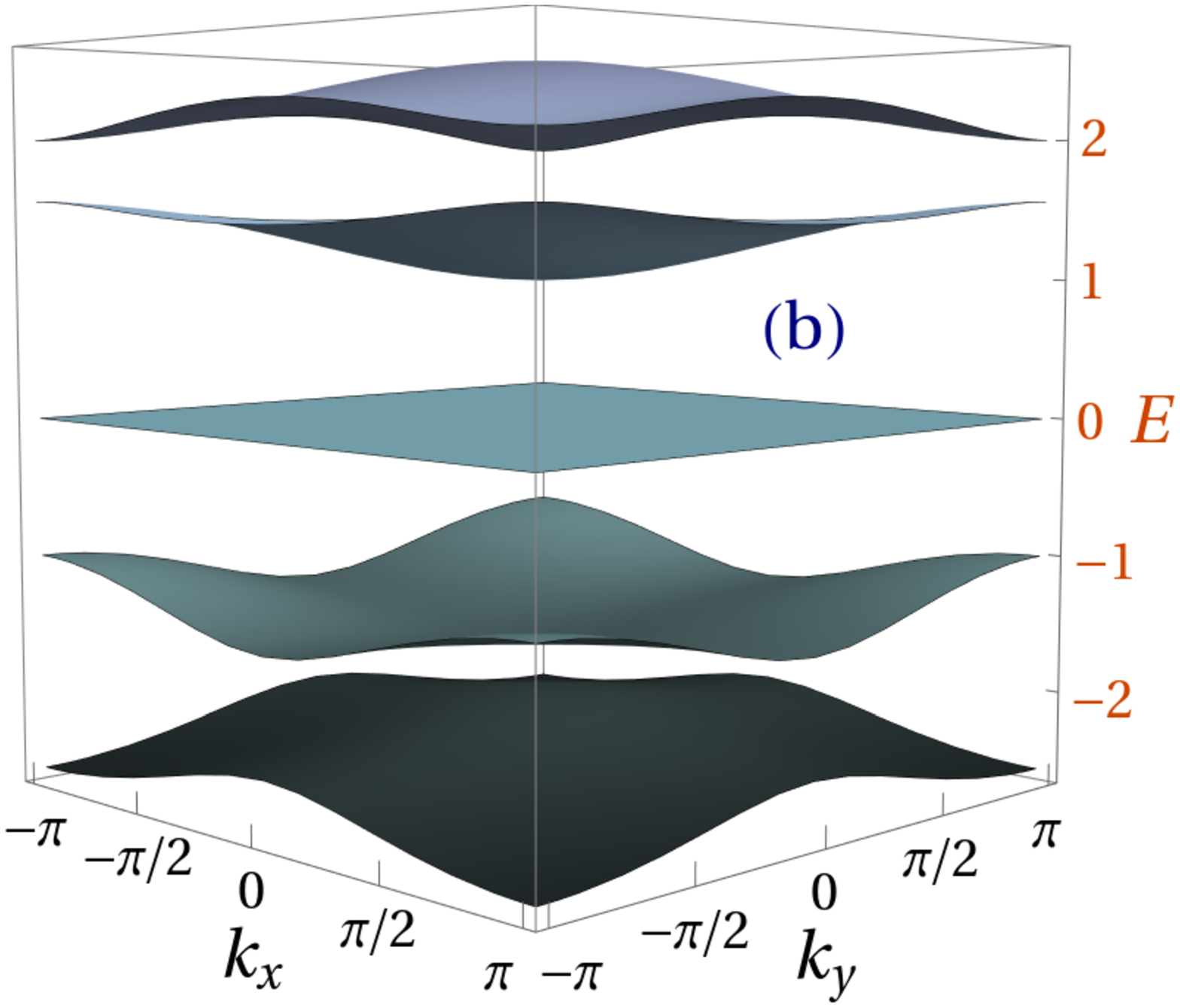}
\caption{Band dispersions for the extended Lieb lattice model in
the presence of intrinsic spin-orbit interaction $\alpha$ (a) 
for $\alpha=0.1$ and (b) for $\alpha=0.5$. Gaps open up 
between all the bands. The values of the hopping amplitudes $t$ 
and $\lambda$ are the same as in Fig.~\ref{fig:FB-without-SO}.}
\label{fig:band-with-ISO}
\end{figure}
We note that $\bm{\mathcal{\hat{H}}_{\textrm{ISO}}}$ does not have a $\mathbf{k}$ dependence since all the possible 
next-nearest-neighbor sites are within the same unit cell for our model (see Fig.~\ref{fig:Lieb-lattice}). The total 
Hamiltonian of the system corresponding to a spin species will now be a combination of Eqs.~\eqref{eq:hamil-k} 
and~\eqref{eq:hamil-iso-mom}. As we tune the ISO coupling strength $\alpha$ to a nonzero value, the band spectrum of 
the system gets gapped out, as shown in Fig.~\ref{fig:band-with-ISO}. In general, as we change the value of $\alpha \neq 0$, 
the completely flat bands in the spectrum get destroyed as displayed in Fig.~\ref{fig:band-with-ISO}(a). However, for 
the special value of $\alpha=0.5$, an isolated completely flat band reappears at the band center with energy $E=0$. 
This case is exhibited in Fig.~\ref{fig:band-with-ISO}(b). The intrinsic spin-orbit coupling parameter $\alpha$ 
plays the role of an effective magnetic field in this tight-binding lattice model. In general, such a magnetic flux destroys 
the existing flat bands in system, but for certain combinations of such a magnetic flux, a completely flat nondispersive band 
reappears in the band structure of the lattice model~\cite{green-prb10, nagaosa-prb00}. Similarly, for the present lattice 
model, in general, for a nonzero finite value of $\alpha$, the existing flat bands in the spectrum get destroyed, and for 
the special value of $\alpha=0.5$, one of the bands becomes completely flat. For this special combination 
of the parameter $\alpha$ along with other system parameters, a destructive quantum interference takes place at the energy 
eigenvalue $E=0$, and the corresponding eigenstates get completely localized forming a flat band at that energy in the band center. 
The corresponding CLS amplitude distribution is shown in Fig.~\ref{fig:amp-distribution-E0}. 
\begin{figure}[ht]
\includegraphics[clip, width=0.8\columnwidth]{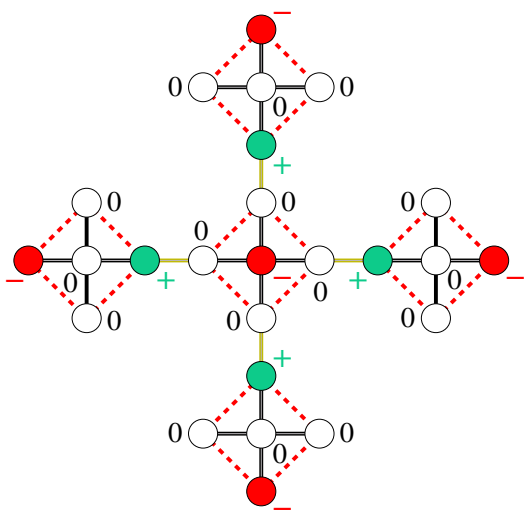}
\caption{Plot of the CLS amplitude distribution corresponding to the flat band 
appearing at energy $E=0$ as shown in Fig.~\ref{fig:band-with-ISO}(b).}
\label{fig:amp-distribution-E0}
\end{figure}
The opening of band gaps in the spectrum is an important 
criterion to identify interesting nontrivial topological phases in the system. Hence, a relevant question to ask is 
whether the resulting gaps for our model are topologically nontrivial. This question is addressed in the next section. 
\section{Identification of the nontrivial topological phase}
\label{topo-properties} 
To gain insight into whether the present lattice model supports nontrivial topological properties, we compute 
the Berry curvatures and the corresponding Chern numbers for all the bands. The Berry curvature of the $n$th band is 
given by~\cite{haldane-prl04, chen-jpcm12} 
\begin{align}
& \mathcal{F}(E_{n},\mathbf{k}) = \nonumber \\
& \hspace{-3mm}\sum_{E_{m} (\neq E_{n})}\hspace{-3.5mm} \dfrac{-2\,\textrm{Im}\Big[ \langle \psi_{n}(\mathbf{k}) | 
\dfrac{\partial \bm{\mathcal{H}}(\mathbf{k})}{\partial k_{x}} | \psi_{m}(\mathbf{k}) \rangle 
\langle \psi_{m}(\mathbf{k}) | 
\dfrac{\partial \bm{\mathcal{H}}(\mathbf{k})}{\partial k_{y}} | \psi_{n}(\mathbf{k}) \rangle \Big]} 
{(E_{n}-E_{m})^2},
\label{eq:BC}
\end{align}
where $\bm{\mathcal{H}}=\bm{\mathcal{H}}_{0} + \bm{\mathcal{H}}_{\textrm{ISO}}$ and $\psi_{n}(\mathbf{k})$ is the $n$th 
eigenstate of $\bm{\mathcal{H}}(\mathbf{k})$ with eigenvalue $E_{n}(\mathbf{k})$. From Eq.~\eqref{eq:BC} we can work out 
the Chern number $\mathcal{C}_{n}$ associated with band $n$ as follows: 
\begin{equation}
\mathcal{C}_{n} = \frac{1}{2 \pi}\int_{1BZ}\mathcal{F}(E_{n},\mathbf{k})d{\mathbf{k}},
\label{eq:Chern}
\end{equation}
where $1BZ$ indicates that the integral is over the first Brillouin zone of the related lattice model. 

It is important to note that, for the Chern number to be well defined, we need to have gaps between all the bands in the spectrum; 
that is, if the two bands touch each other at any point, then the Chern number will be ill-defined. To resolve the issue of the spin 
degeneracy of the bands for our model, we project the system into a single spin component and then calculate the Chern numbers 
for the spin-up ($\uparrow$) and spin-down ($\downarrow$) bands separately as $\mathcal{C}_{n}^{\uparrow}$ and 
$\mathcal{C}_{n}^{\downarrow}$, respectively. 
\begin{figure}[ht]
\includegraphics[clip, width=0.49\columnwidth]{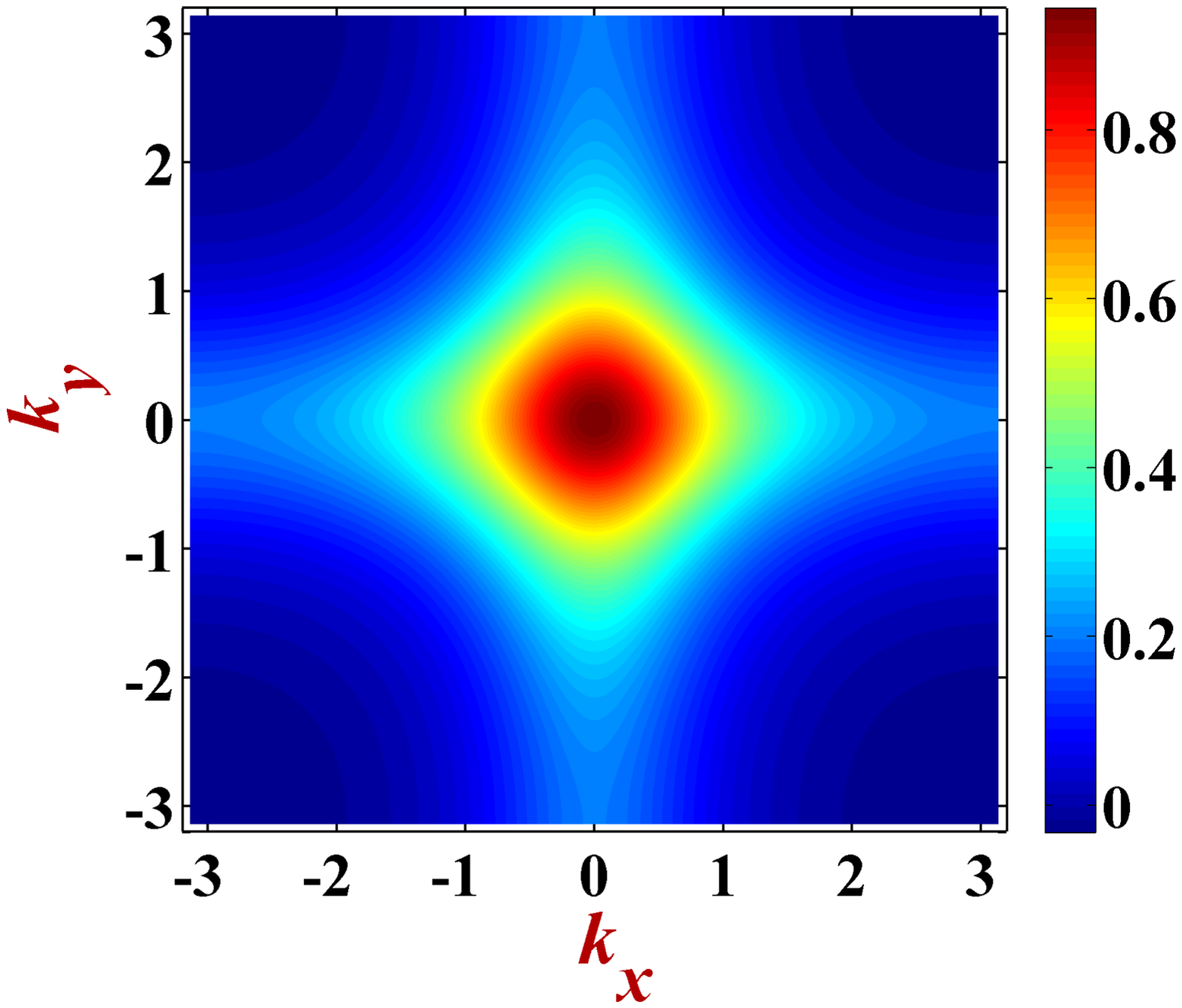}
\includegraphics[clip, width=0.49\columnwidth]{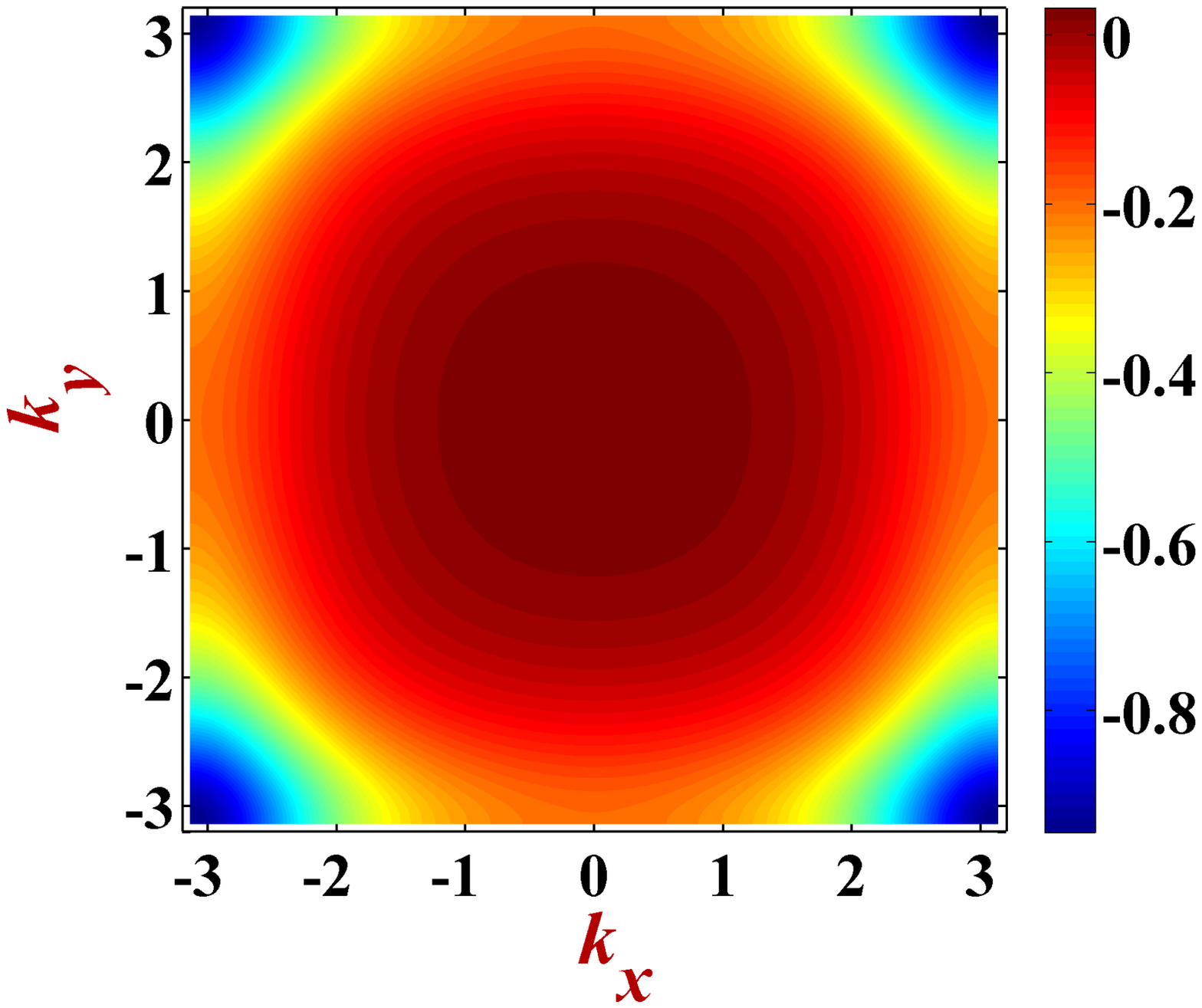}
\caption{Distribution of Berry curvature in the momentum space 
corresponding to the topologically nontrivial bands (for the 
spin-up ($\uparrow$) case) with integer Chern numbers for 
$\alpha=0.5$. The left panel shows the lowest band ($n=1$) with 
$\mathcal{C}_{n=1}^{\uparrow}=1$, and the right panel corresponds 
to the highest band ($n=5$) with $\mathcal{C}_{n=5}^{\uparrow}=-1$. 
The values of the hopping amplitudes are taken to be $t=1$ and 
$\lambda=1$, respectively.}
\label{fig:BC_alpha_0-5}
\end{figure}
The topological invariant for such spin-dependent two-dimensional lattice models is defined as spin Chern number 
$\mathcal{C}^{s}_{n} = \mathcal{C}_{n}^{\uparrow} - \mathcal{C}_{n}^{\downarrow}$~\cite{morais-smith-prb12, 
haldane-prl06, zhang-pla16}. As TRS is preserved in the system, the total charge Chern number 
$\mathcal{C}^{c}_{n} = \mathcal{C}_{n}^{\uparrow} + \mathcal{C}_{n}^{\downarrow}$~\cite{haldane-prl06} is zero for 
this model. It follows that $\mathcal{C}_{n}^{\downarrow} = -\mathcal{C}_{n}^{\uparrow}$ leading to 
$\mathcal{C}^{s}_{n} = 2\mathcal{C}_{n}^{\uparrow}$.  

Following the above prescription, we have computed the Chern numbers $\mathcal{C}_{n}^{\uparrow}$ and 
$\mathcal{C}_{n}^{\downarrow}$ in the presence of a nonzero value of $\alpha=0.5$. We have found the corresponding values 
to be $\mathcal{C}_{n}^{\uparrow} = (1,0,0,0,-1)$ and $\mathcal{C}_{n}^{\downarrow} = (-1,0,0,0,1)$, respectively. In 
these expressions, the values from left to right correspond to going from the lowest to highest bands. It is clearly 
evident from the above values that the lowest and highest bands in the spectrum are topologically nontrivial with 
nonzero values of the Chern numbers, while the bands in the middle appear to be topologically trivial with zero Chern 
numbers. We note that, for $\alpha=0.5$, there is a completely flat band in the middle of the spectrum 
[see Fig.~\ref{fig:band-with-ISO}(b)]. Hence, the fact that a completely flat band cannot have a nonzero Chern number 
is also reflected by our result of the Chern numbers. This is true because we cannot have completely flat 
band, nonzero Chern number, and short-range local hopping simultaneously in a single tight-binding model --- it is possible to 
satisfy any two of these three conditions simultaneously~\cite{chen-jpa14}. We have exhibited the distribution of the Berry 
curvatures in the momentum space corresponding to the topologically nontrivial bands for the spin-up ($\uparrow$) case in 
Fig.~\ref{fig:BC_alpha_0-5}. We can easily find the marked features appearing in the distribution of the Berry curvatures for such 
topological Chern bands as evident in Fig.~\ref{fig:BC_alpha_0-5}.
\begin{figure}[ht]
\includegraphics[clip, width=0.49\columnwidth]{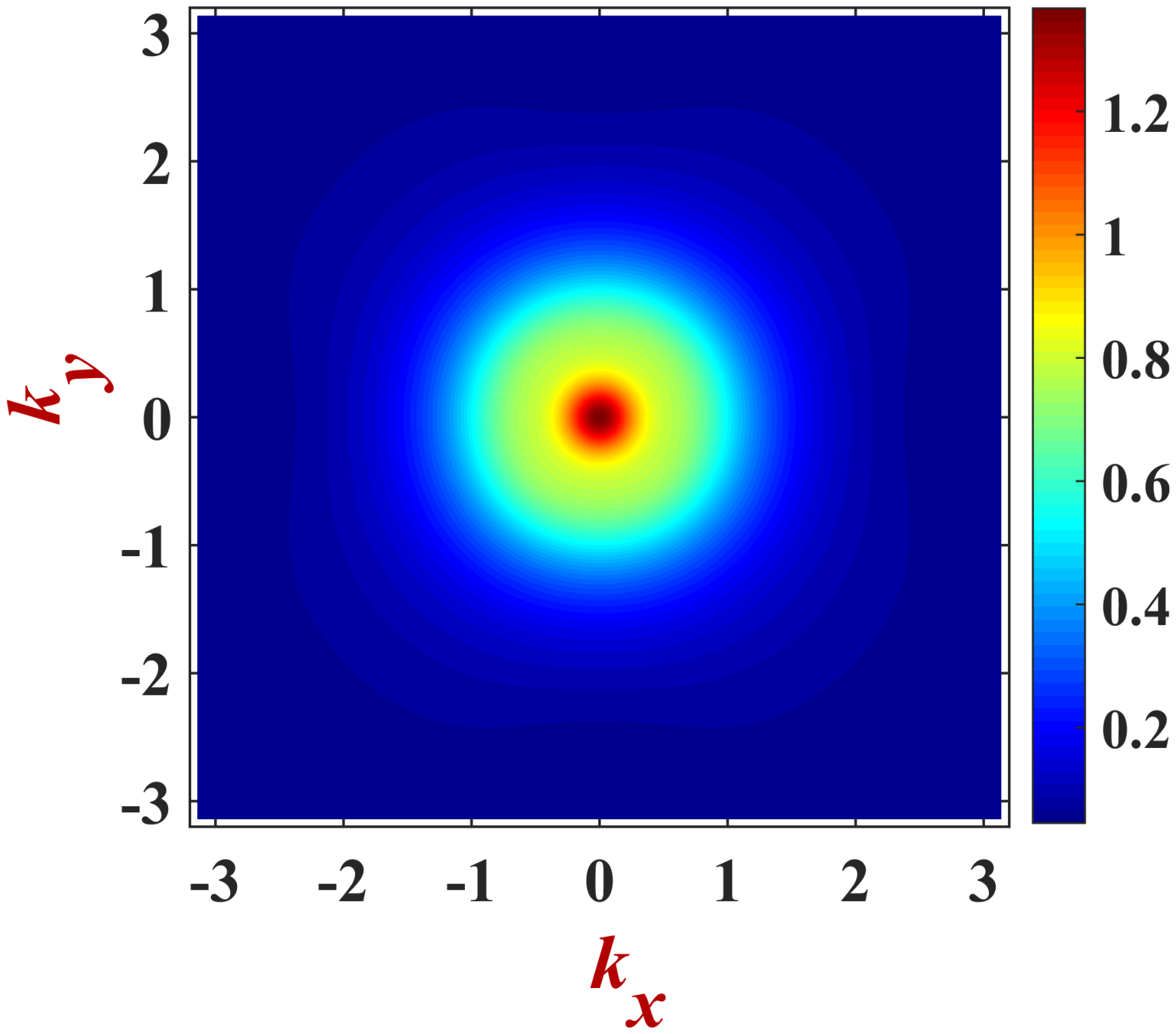}
\includegraphics[clip, width=0.49\columnwidth]{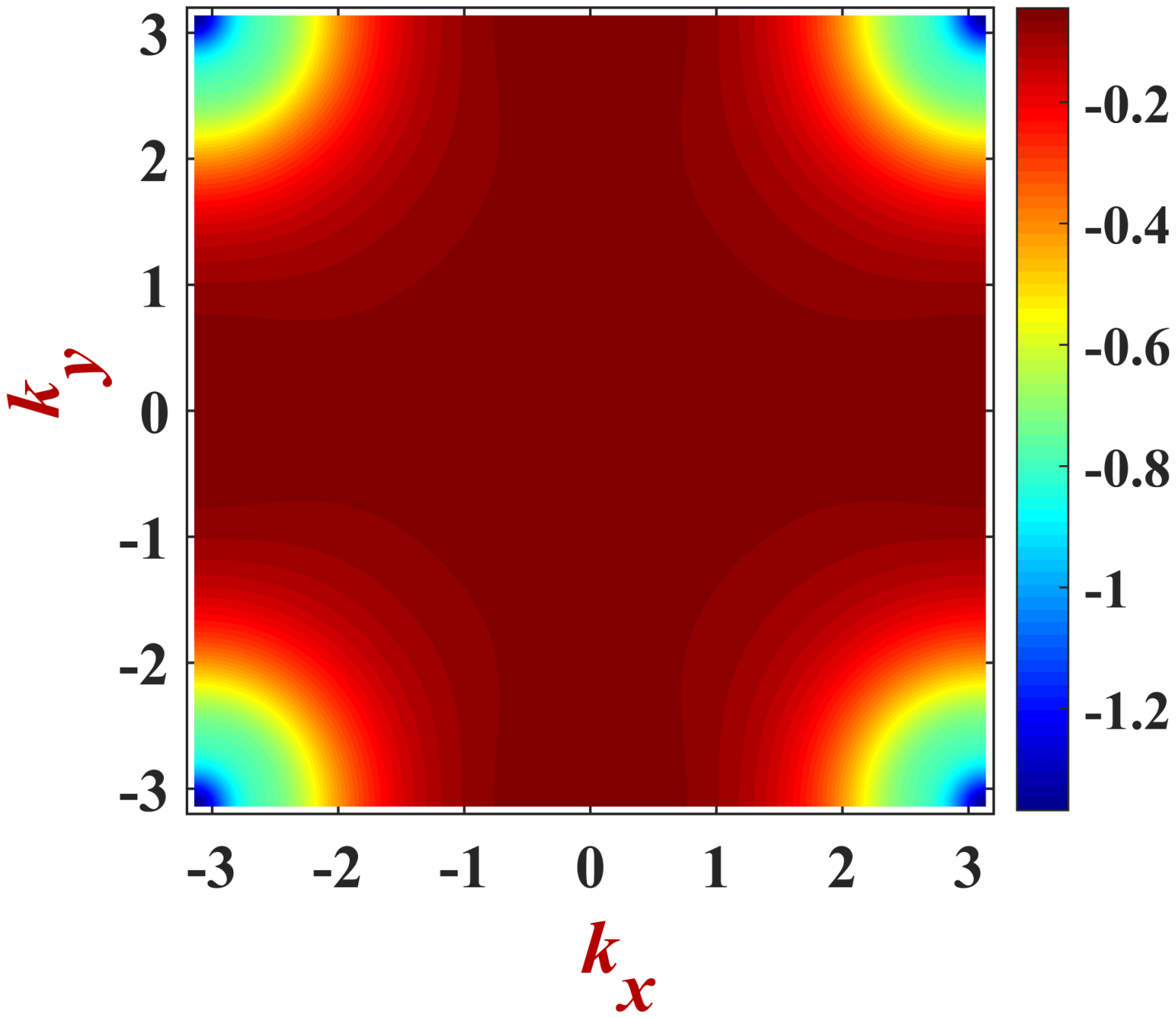}
\caption{Distribution of Berry curvature in the momentum space 
corresponding to the topologically nontrivial bands (for the 
spin-up ($\uparrow$) case) with integer Chern numbers for 
$\alpha=0.1$. The left panel shows the second band ($n=2$) with 
$\mathcal{C}_{n=2}^{\uparrow}=1$, and the right panel corresponds 
to the fourth band ($n=4$) with $\mathcal{C}_{n=4}^{\uparrow}=-1$. 
The values of the hopping amplitudes are the same as in 
Fig.~\ref{fig:BC_alpha_0-5}.}
\label{fig:BC_alpha_0-1}
\end{figure}

The values of these topological measures $\mathcal{C}_{n}^{\uparrow}$ $(\mathcal{C}_{n}^{\downarrow})$ remain unchanged 
until the energy gap collapses. As we change the value of $\alpha$, there will be a certain point where one might observe a 
topological phase transition. At such a point, the values of the Chern numbers change for different bands. For our model we 
witness such a phenomenon; for example, we obtain $\mathcal{C}_{n}^{\uparrow} = (0,1,0,-1,0)$ and 
$\mathcal{C}_{n}^{\downarrow} = (0,-1,0,1,0)$, corresponding to the band spectrum in 
Fig.~\ref{fig:band-with-ISO}(a) with $\alpha=0.1$. So there is a distinct 
topological phase transition in the system as reflected by the changes in the values of Chern numbers for different bands. 
Such a topological phase transition is always associated with an energy gap closing and then reopening around the transition 
point. The Berry curvature distributions corresponding to these Chern bands for $\alpha=0.1$ are shown in Fig.~\ref{fig:BC_alpha_0-1}. 
Interestingly, it should be noted that, for $\alpha=0.1$, the bands ($n=2$ and $n=4$) which pick up the nonzero Chern 
numbers are almost flat. Such nearly flat bands with nonzero Chern numbers may be treated analogously to the flat degenerate 
Landau levels appearing in a continuum model in the presence of a real magnetic field. Hence, this might lead to an interesting 
possibility of realizing fractional quantum spin Hall physics in this lattice model when the interaction between particles is 
taken into account. The value of the spin Chern number is a measure of the spin Hall conductivity in the system, and both the 
quantities are related in the following way~\cite{morais-smith-prb12, haldane-prl06}:  
\begin{equation}
\sigma_{\textrm{SHC}}=\dfrac{e}{4\pi}\sum_{n^{\prime}}
\mathcal{C}^{s}_{n^{\prime}},
\label{eq:SHC}
\end{equation}
where the summation is over all the filled bands $n^{\prime}$ with energy $E_{n^{\prime}}<E_{F}$, with $E_{F}$ being the Fermi 
energy level. Thus, in our model for different band fillings we will get nonzero values of $\sigma_{\textrm{SHC}}$. As 
the spin Hall conductivity $\sigma_{\textrm{SHC}}$ is connected to the $\mathbb{Z}_{2}$ topological index 
$\nu=\sigma_{\textrm{SHC}}/2\ \textrm{mod}\ 2$~\cite{morais-smith-prb12}, we may conclude that our lattice model can, indeed, be 
a potential host for a topological insulator in the presence of intrinsic spin-orbit coupling. 

\begin{figure}[ht]
\includegraphics[clip, width=0.49\columnwidth]{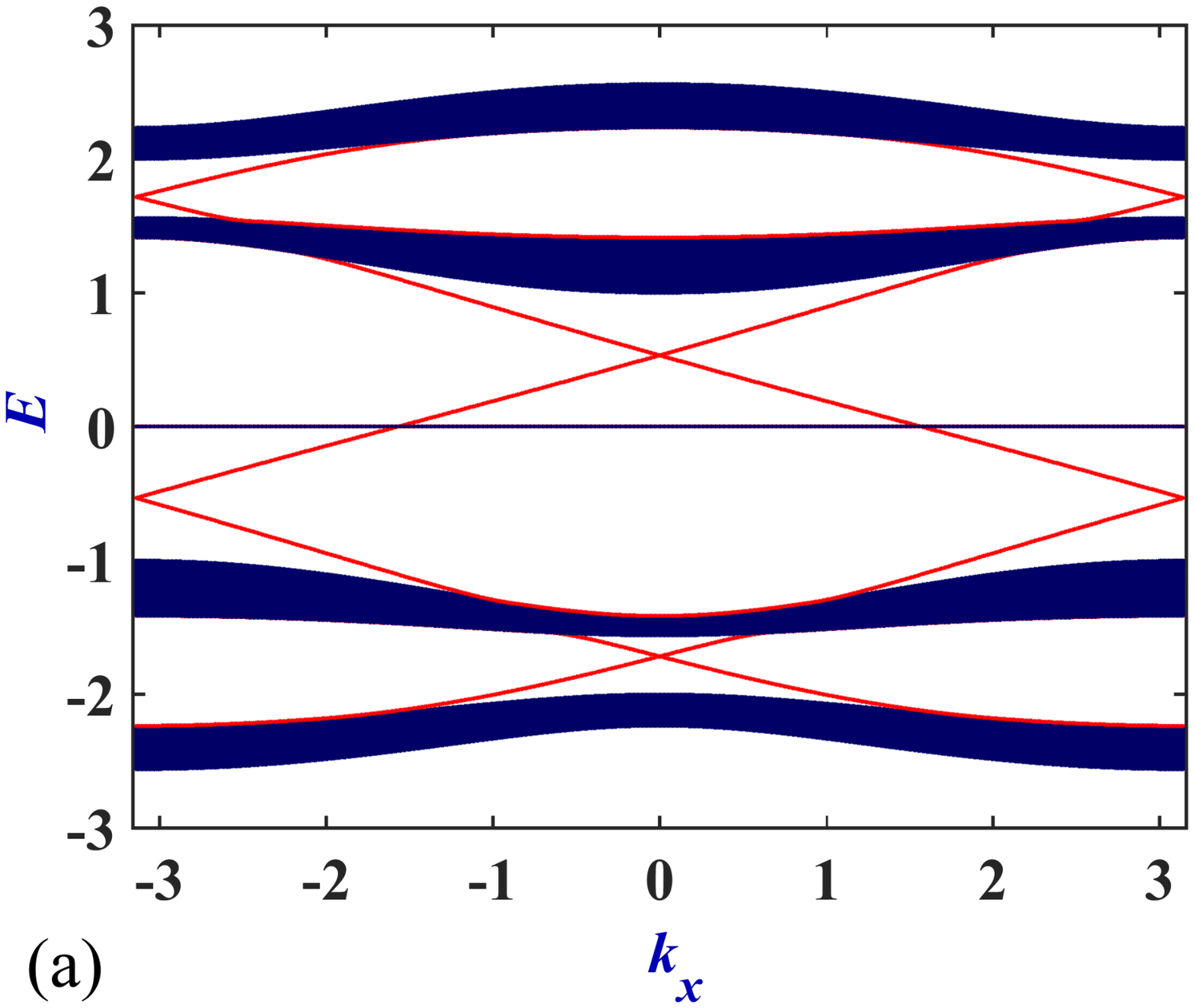}
\includegraphics[clip, width=0.49\columnwidth]{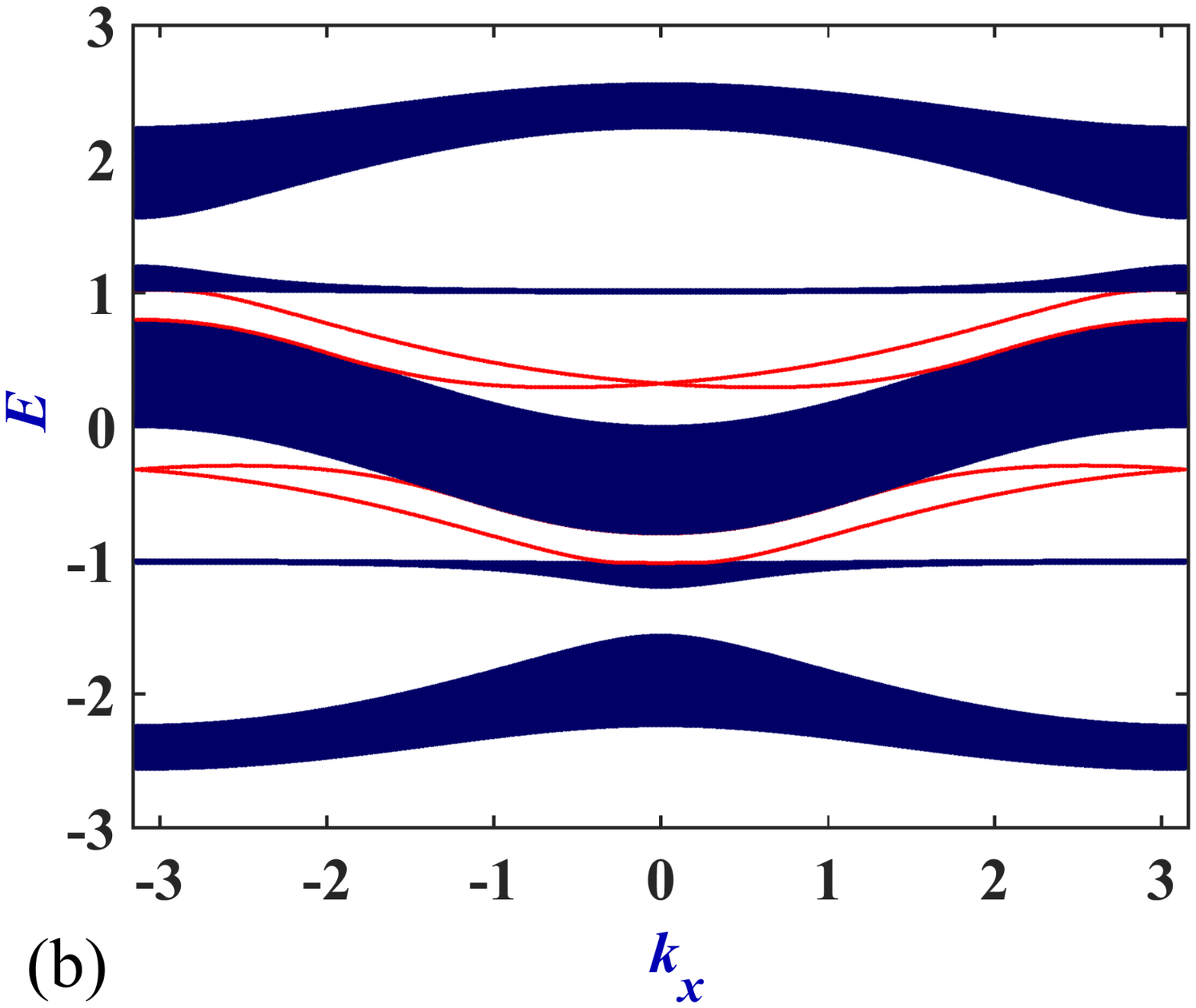}
\caption{Distribution of the edge states for the extended Lieb 
lattice geometry with (a) $\alpha=0.5$ and (b) $\alpha=0.1$. 
We have taken a periodic boundary condition 
along the $x$ direction, and the number of unit cells in the 
$y$ direction is taken to be 100. The bulk bands are shown in 
dark blue, and the edge states are depicted by red 
lines. The values of the hopping amplitudes are taken to be 
$t=1$ and $\lambda=1$, respectively.}
\label{fig:Edge-states}
\end{figure}
To further demonstrate the topological properties of the system, we compute the edge states~\cite{kane-mele-prl05} for 
this lattice model. We use a standard recipe to compute the edge states, which is to effectively place the system on a 
cylinder; that is, we consider the periodic boundary condition in one direction of the lattice geometry and truncate it to a 
finite size in the other direction. We construct the Hamiltonian for such a truncated strip with the periodic boundary in one 
direction and diagonalize it to get the edge states lying inside the gaps between the bulk bands. We show the results 
in Fig.~\ref{fig:Edge-states} for $\alpha=0.5$ and $\alpha=0.1$. Here, we have considered a periodic boundary 
condition along the $x$ direction and an open boundary along the $y$ direction with $100$ unit cells. We can clearly observe 
the presence of the edge states (indicated by red lines) inside the bulk band gaps in Fig.~\ref{fig:Edge-states}. The 
nonzero values of the Chern numbers manifest in the presence of these edge states for this lattice model. The number of 
edge states appearing inside the bulk gaps in the band structure is related to the sum of the Chern numbers associated with 
the filled bands via the following relation~\cite{cm-smith-prb12}: 
\begin{equation}
{\mathcal{N}}_{p}=\sum_{n\in\:\textrm{filled bands}}
\mathcal{C}_{n},
\label{eq:edgestates}
\end{equation}
where ${\mathcal{N}}_{p}$ is the number of edge states in the $p$th bulk gap. This is known as the bulk-boundary correspondence since 
it connects the properties of the system in the surface and in the bulk. 
\section{Possible application of the model}
\label{expt-setup}
Experimental realization of the interesting topological phases found in this lattice geometry using ultracold atoms in 
an analogous optical lattice setup can be a compelling task to execute. Over the last couple of years, ultracold atoms 
trapped in an optical lattice hasve become an ideal playground to probe various condensed matter phenomena ranging from 
quasi-free to strongly correlated. These artificial crystals of light created by two or more interfering 
laser beams offer a unique setup with full control over a wide range of system parameters, such as hopping amplitudes, 
interaction strength, potential depth etc.~\cite{bloch-rpm08}. Such a high degree of control and fine tunability of the 
parameters allow us to access a regime which is otherwise  unreachable in real condensed-matter systems. 

In cold-atom systems, an atom's internal state plays the role of the spin state in the absence of the real spin~\cite{zhang-pla16, liu-pra10}. 
This adds an important advantage to the measurement process --- one can directly evaluate the spin Chern number by optically 
measuring the internal state of the atoms. In these kinds of systems, there is a connection between the spin Chern number and 
the spin-atomic density~\cite{liu-pra10}. This allows one to detect the topological properties of the system by measuring the 
spin-atomic density by using standard density-profile measurement techniques employed for cold-atomic systems~\cite{liu-pra10, 
shao-prl08}. An alternative way to detect the topological properties in cold-atom systems is to image the corresponding 
edge states~\cite{hofstetter-pra12}. However, in an optical lattice, as the ultracold atoms are confined in a harmonic trap 
which smoothly varies in space, there is a lack of sharp edges in the system. Such soft edges in the system may modify and 
destroy the usual ordered band structure of the system consisting of bulk bands and edge states in the gaps. Nevertheless, it 
has been shown that, even for the system with soft edges, the topological invariants remain rigid~\cite{hofstetter-pra12, goldman-prl12}. 
\begin{figure}[ht]
\includegraphics[clip, width=0.9\columnwidth]{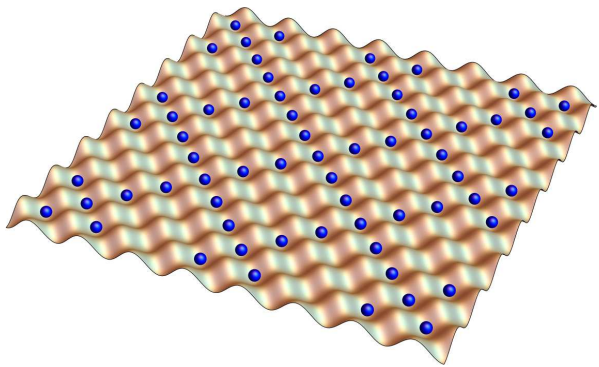}
\caption{Schematic representation of an extended Lieb optical lattice 
model. Ultracold atoms may be trapped deterministically in a 
two-dimensional periodic potential pattern formed by interfering 
laser beams to create such an artifact mimicking the lattice 
geometry depicted in Fig.~\ref{fig:Lieb-lattice}.}
\label{fig:optical-lattice}
\end{figure}

In this context, we propose that given the simple geometrical structure of the system, the present lattice model might be 
translated into an analogous optical lattice setup and the corresponding physical properties can be easily measured using 
the methods mentioned above. Spin-orbit coupling, being an atom's intrinsic property, cannot be varied in a real 
condensed-matter system. However, ultracold atoms in an optical lattice are a promising platform in this regard. Intrinsic 
spin-orbit coupling which is a complex next-nearest-neighbor hopping in other words, can be tuned in the cold-atom setup by 
circular periodic modulation of the lattice~\cite{esslinger-nature14}. Nevertheless, in the present case, the phase of the 
next-nearest-neighbor hopping for two spin projections (spin up and spin down) should be equal but opposite, such that 
the entire system remains time reversal symmetric. This condition can be achieved by trapping the ultracold atoms in 
spin-dependent lattice (i.e., one for spin-up particle and the other one, superimposed, for spin-down particles) and then 
shaking these two lattices circularly but with opposite orientation for two spin directions. We note that a 
time-reversal-symmetric topological insulator was already realized in a cold-atom setup~\cite{goldman-prl10, monica-prl13}.
So the above method may enable us to produce the interesting topological features found for the present lattice model in 
a possible real-life experimental setup. To create such an analogous optical lattice corresponding to 
Fig.~\ref{fig:Lieb-lattice}, one has to generate a periodic potential well in two dimensions using suitable combinations 
of \textit{sine} and \textit{cosine} potential functions~\cite{goldman-pra11} and then trap the neutral ultracold atoms 
inside such a potential pattern in a deterministic way to form the desired lattice pattern. Such a complex architecture of 
an extended Lieb optical lattice model is envisioned schematically in Fig.~\ref{fig:optical-lattice}.  
\section{Concluding remarks and future outlook}
\label{conclu}
In conclusion, we have studied the emergence of interesting flat-band physics and nontrivial topological 
properties in an extended Lieb lattice model with five atomic sites per unit cell. We have shown that, this lattice geometry 
offers multiple numbers of completely flat nondispersive bands in its band spectrum only with a nearest-neighbor hopping model. 
These completely flat bands are accompanied by other gapped and nongapped dispersive bands in the band structure. This is in 
marked contrast to the band structure of a conventional three-site unit cell Lieb lattice model, where one encounters a single 
flat band sandwiched between two dispersive bands which form a Dirac-cone-like structure. Such a Dirac cone does not appear for this 
extended Lieb lattice model, and also we have more than one flat band in the spectrum. 

With the inclusion of an intrinsic spin-orbit interaction term in the corresponding tight-binding Hamiltonian, band gaps 
open up in the system for this lattice model. Such intrinsic spin-orbit coupling acts like a complex next-nearest-neighbor 
hopping amplitude for this model, and it plays the role of an effective magnetic field for the system. However, an 
important feature is that it does not break the time-reversal symmetry of the system. In the presence of such time-reversal 
symmetry in the system, this model allows for the manifestation of nontrivial topological characters of certain bands in 
the gapped spectrum. This is explicitly verified by computation of the spin Chern numbers for different bands appearing in 
the band spectrum. We have obtained nonzero integer values of the spin Chern numbers, suggesting the appearance of 
interesting nontrivial topological quantum spin Hall phases for this lattice model. In addition to that, for certain 
values of the intrinsic spin-orbit interaction strength, we have identified the appearance of almost flat bands in the 
spectrum with nonzero integer values of the Chern numbers. This opens up the interesting possibility of exploring the 
fractional quantum spin Hall effect in this lattice model when the interaction among the particles is taken into 
consideration. Finally, we propose that this model and the corresponding results might be tested and accomplished 
experimentally using an analogous optical lattice setup with ultracold atomic condensates. 

In the present study, we have explored the topological properties of a variation of edge-centered square lattice 
model. It might be really interesting to carry forward this idea to other classes of edge-centered square lattice models and to 
examine if we can also have interesting topological properties emerging out of such systems. Another significant futuristic 
direction of the present work could be to investigate the effect of Rashba spin-orbit interaction or Zeeman field or the 
combination of both on the flat bands and topological properties of such different classes of extended Lieb lattice models. 
The study of photonic flat-band localization and nonlinear effects in an extended photonic Lieb lattice model could also be 
an interesting future direction of the present study.
\begin{acknowledgments}
We are indebted to N. Goldman for sharing his valuable idea on the experimental realization of intrinsic spin-orbit 
coupling using cold atoms in optical lattice. 
B.P. would like to acknowledge the financial support provided through a Kreitman postdoctoral fellowship. 
This work is also partially supported through funding by the Israel Science Foundation (ISF), by the infrastructure 
program of Israel Ministry of Science and Technology under Contract No. 3-11173, and by the Pazy Foundation.
\end{acknowledgments}


\begin{thebibliography}{99}

\bibitem{stern-aop08} A. Stern,
\href{https://doi.org/10.1016/j.aop.2007.10.008}
{Ann. Phys. \textbf{323}, 204 (2008)}. 

\bibitem{kane-prl08} L. Fu and C. L. Kane, 
\href{https://doi.org/10.1103/PhysRevLett.100.096407}
{Phys. Rev. Lett. \textbf{100}, 096407 (2008)}. 

\bibitem{alicea-rpp12} J. Alicea, 
\href{https://doi.org/10.1088/0034-4885/75/7/076501}
{Rep. Prog. Phys. \textbf{75}, 076501 (2012)}. 

\bibitem{nandkishore-arcmp19} R. M. Nandkishore and M. Hermele, 
\href{https://doi.org/10.1146/annurev-conmatphys-031218-013604}
{Annu. Rev. Condens. Matter Phys. \textbf{10}, 295 (2019)}. 

\bibitem{nayek-prl05} S. Das Sarma, M. Freedman, and C. Nayak, 
\href{https://doi.org/10.1103/PhysRevLett.94.166802}
{Phys. Rev. Lett. \textbf{94}, 166802 (2005)}. 

\bibitem{kane-mele-prl05} C. L. Kane and E. J. Mele, 
\href{https://doi.org/10.1103/PhysRevLett.95.146802}
{Phys. Rev. Lett. \textbf{95}, 146802 (2005)}; 
C. L. Kane and E. J. Mele, 
\href{https://doi.org/10.1103/PhysRevLett.95.226801}
{Phys. Rev. Lett. \textbf{95}, 226801 (2005)}. 

\bibitem{bernevig-prl06} B. A. Bernevig and S.-C. Zhang,
\href{https://doi.org/10.1103/PhysRevLett.96.106802}
{Phys. Rev. Lett. \textbf{96}, 106802 (2006)}. 

\bibitem{molenkamp-sci07} M. K\"{o}nig, S. Wiedmann, C. Br\"{u}ne, A. Roth, 
H. Buhmann, L. W. Molenkamp, X.-L. Qi, and S.-C. Zhang,
\href{https://doi.org/10.1126/science.1148047}
{Science \textbf{318}, 766 (2007)}. 

\bibitem{wang-bookchapter16} K. L. Wang, M. Lang, and X. Kou, 
\href{https://doi.org/10.1007/978-94-007-6892-5_56}
{Spintronics of Topological Insulators}, in \textit{Handbook of Spintronics}, 
edited by Y. Xu, D. Awschalom, and J. Nitta (Springer, Dordrecht, 2016), pp. 431-462. 

\bibitem{wang-spin16} Y. Fan and K. L. Wang, 
\href{https://doi.org/10.1142/S2010324716400014}
{SPIN \textbf{06}, 1640001 (2016)}. 

\bibitem{moore-nature10} J. E. Moore,
\href{https://doi.org/10.1038/nature08916}
{Nature (London) \textbf{464}, 194 (2010)}. 

\bibitem{hasan-kane-rmp10} M. Z. Hasan and C. L. Kane, 
\href{https://doi.org/10.1103/RevModPhys.82.3045}
{Rev. Mod. Phys. \textbf{82}, 3045 (2010)}.  

\bibitem{zhang-rmp11} X.-L.  Qi  and  S.-C.  Zhang, 
\href{https://doi.org/10.1103/RevModPhys.83.1057}
{Rev. Mod. Phys. \textbf{83}, 1057 (2011)}.  

\bibitem{kane-prl07} L. Fu, C. L. Kane, and E. J. Mele, 
\href{https://doi.org/10.1103/PhysRevLett.98.106803}
{Phys. Rev. Lett. \textbf{98}, 106803 (2007)}. 

\bibitem{franz-prb09} H.-M. Guo and M. Franz, 
\href{https://doi.org/10.1103/PhysRevB.80.113102}
{Phys. Rev. B \textbf{80}, 113102 (2009)}. 

\bibitem{kaisun-prl09} K. Sun, H. Yao, E. Fradkin, and S. A. Kivelson, 
\href{https://doi.org/10.1103/PhysRevLett.103.046811}
{Phys. Rev. Lett. \textbf{103}, 046811 (2009)}. 

\bibitem{fiete-prb10} A. R\"{u}egg, J. Wen, and G. A. Fiete, 
\href{https://doi.org/10.1103/PhysRevB.81.205115}
{Phys. Rev. B \textbf{81}, 205115 (2010)}. 

\bibitem{mielke-jpa91} A. Mielke,
\href{https://doi.org/10.1088/0305-4470/24/14/018}
{J. Phys. A: Math. Gen. \textbf{24}, 3311 (1991)}. 

\bibitem{das-sharma-prl11} K. Sun, Z. Gu, H. Katsura, and S. Das Sarma, 
\href{https://doi.org/10.1103/PhysRevLett.106.236803}
{Phys. Rev. Lett. \textbf {106}, 236803 (2011)}. 

\bibitem{tang-prl11} E. Tang, J.-W. Mei, and X.-G. Wen, 
\href{https://doi.org/10.1103/PhysRevLett.106.236802}
{Phys. Rev. Lett. \textbf {106}, 236802 (2011)}. 

\bibitem{titus-prl11} T. Neupert, L. Santos, C. Chamon, and C. Mudry, 
\href{https://doi.org/10.1103/PhysRevLett.106.236804}
{Phys. Rev. Lett. \textbf{106}, 236804 (2011)}. 

\bibitem{flach-prb13} D. Leykam, S. Flach, O. Bahat-Treidel, and A. S. Desyatnikov, 
\href{https://doi.org/10.1103/PhysRevB.88.224203}
{Phys. Rev. B  \textbf{88}, 224203 (2013)}. 

\bibitem{flach-epl14} S. Flach, D. Leykam, J. D. Bodyfelt, P. Matthies, and A. S. Desyatnikov, 
\href{https://doi.org/10.1209/0295-5075/105/30001}
{Europhys. Lett.  \textbf{105}, 30001 (2014)}. 

\bibitem{flach-prl14} J. D. Bodyfelt, D. Leykam, C. Danieli, X. Yu, and S. Flach, 
\href{https://doi.org/10.1103/PhysRevLett.113.236403}
{Phys. Rev. Lett. \textbf{113}, 236403 (2014)}. 

\bibitem{flach-prb15} C. Danieli, J. D. Bodyfelt, and S. Flach, 
\href{https://doi.org/10.1103/PhysRevB.91.235134}
{Phys. Rev. B  \textbf{91}, 235134 (2015)}. 

\bibitem{flach-prl16} R. Khomeriki and S. Flach,
\href{https://doi.org/10.1103/PhysRevLett.116.245301}
{Phys. Rev. Lett. \textbf{116}, 245301 (2016)}. 

\bibitem{vicencio-pra17} S. Rojas-Rojas, L. Morales-Inostroza, R. A. Vicencio, and A. Delgado, 
\href{https://doi.org/10.1103/PhysRevA.96.043803}
{Phys. Rev. A \textbf{96}, 043803 (2017)}. 

\bibitem{flach-prb17} W. Maimaiti, A. Andreanov, H. C. Park, O. Gendelman, and S. Flach, 
\href{https://doi.org/10.1103/PhysRevB.95.115135}
{Phys. Rev. B  \textbf{95}, 115135 (2017)}. 

\bibitem{ajith-prb18} A. R. Kolovsky, A. Ramachandran, and S. Flach
\href{https://doi.org/10.1103/PhysRevB.97.045120}
{Phys. Rev. B \textbf{97}, 045120 (2018)}. 

\bibitem{ajith-prb17} A. Ramachandran,  A. Andreanov, and S. Flach,
\href{https://doi.org/10.1103/PhysRevB.96.161104}
{Phys. Rev. B \textbf{96}, 161104(R) (2017)}. 

\bibitem{biplab-prb18} B. Pal and K. Saha, 
\href{https://doi.org/10.1103/PhysRevB.97.195101}
{Phys. Rev. B \textbf{97}, 195101 (2018)}. 

\bibitem{pal-prb18} B. Pal, 
\href{https://doi.org/10.1103/PhysRevB.98.245116}
{Phys. Rev. B \textbf{98}, 245116 (2018)}. 

\bibitem{jun-won-prb19} J.-W. Rhim and B.-J. Yang, 
\href{https://doi.org/10.1103/PhysRevB.99.045107}
{Phys. Rev. B \textbf{99}, 045107 (2019)}. 

\bibitem{nandy-pla19} A. Nandy and A. Mukherjee, 
\href{https://doi.org/10.1016/j.physleta.2019.04.035}
{Phys. Lett. A \textbf{383}, 2318 (2019)}. 

\bibitem{goda-prl06} M. Goda, S. Nishino, and H. Matsuda, 
\href{https://doi.org/10.1103/PhysRevLett.96.126401}
{Phys. Rev. Lett. \textbf{96}, 126401 (2006)}. 

\bibitem{shukla-prb10} J. T. Chalker, T. S. Pickles, and P. Shukla, 
\href{https://doi.org/10.1103/PhysRevB.82.104209}
{Phys. Rev. B \textbf{82}, 104209 (2010)}. 

\bibitem{tasaki-prl92} H. Tasaki, 
\href{https://doi.org/10.1103/PhysRevLett.69.1608}
{Phys. Rev. Lett. \textbf{69}, 1608 (1992)}. 

\bibitem{richter-prl12} M. Maksymenko, A. Honecker, R. Moessner, J. Richter, and O. Derzhko, 
\href{https://doi.org/10.1103/PhysRevLett.109.096404}
{Phys. Rev. Lett. \textbf{109}, 096404 (2012)}. 

\bibitem{heikkila-prb16} V. J. Kauppila, F. Aikebaier, and T. T. Heikkil\"{a}, 
\href{https://doi.org/10.1103/PhysRevB.93.214505}
{Phys. Rev. B \textbf{93}, 214505 (2016)}. 

\bibitem{torma-natcom15} S. Peotta and P. T\"{o}rm\"{a}, 
\href{https://doi.org/10.1038/ncomms9944}{Nat. Commun. \textbf{6}, 8944 (2015)}. 

\bibitem{torma-prl16} A. Julku, S. Peotta, T. I. Vanhala, D.-H. Kim, and P. T\"{o}rm\"{a}, 
\href{https://doi.org/10.1103/PhysRevLett.117.045303}
{Phys. Rev. Lett. \textbf{117}, 045303 (2016)}. 

\bibitem{montambaux-prl18} G. Montambaux, L.-K. Lim, J.-N. Fuchs, and F. Pi\'{e}chon, 
\href{https://doi.org/10.1103/PhysRevLett.121.256402}
{Phys. Rev. Lett. \textbf{121}, 256402 (2018)}. 

\bibitem{jiang-prb19} W. Jiang, M. Kang, H. Huang, H. Xu, T. Low, and F. Liu, 
\href{https://doi.org/10.1103/PhysRevB.99.125131}
{Phys. Rev. B \textbf{99}, 125131 (2019)}. 

\bibitem{montambaux-arxiv19} L.-K. Lim, J.-N. Fuchs, F. Pi\'{e}chon, and G. Montambaux, 
\href{https://arxiv.org/abs/1906.12002#}
{arXiv: 1906.12002}. 

\bibitem{franz-prb12} C. Weeks and M. Franz, 
\href{https://doi.org/10.1103/PhysRevB.85.041104}
{Phys. Rev. B \textbf{85}, 041104(R) (2012)}.  

\bibitem{liu-prl12} Z. Liu, E. J. Bergholtz, H. Fan, and A. M. L\"{a}uchli, 
\href{https://doi.org/10.1103/PhysRevLett.109.186805}
{Phys. Rev. Lett. \textbf{109}, 186805 (2012)}.  

\bibitem{mukherjee-prl15}  S. Mukherjee, A. Spracklen, D. Choudhury, N. Goldman, 
P. \"{O}hberg, E. Andersson, and R. R. Thomson, 
\href{https://doi.org/10.1103/PhysRevLett.114.245504}
{Phys. Rev. Lett.  \textbf{114}, 245504 (2015)}. 

\bibitem{vicencio-prl15} R. A. Vicencio, C. Cantillano, L. Morales-Inostroza, 
B. Real, C. Mej\'{i}a-Cort\'{e}s, S. Weimann, A. Szameit, and M. I. Molina, 
\href{https://doi.org/10.1103/PhysRevLett.114.245503}
{Phys. Rev. Lett. \textbf{114}, 245503 (2015)}. 

\bibitem{baba-nat-photon08} T. Baba, 
\href{https://doi.org/10.1038/nphoton.2008.146}
{Nat. Photonics \textbf{2}, 465 (2008)}.

\bibitem{mukherjee-ol15} S. Mukherjee and R. R. Thomson, 
\href{https://doi.org/10.1364/OL.40.005443}
{Opt. Lett. \textbf{40}, 5443 (2015)}. 

\bibitem{longhi-ol14} S. Longhi, 
\href{https://doi.org/10.1364/OL.39.005892} 
{Opt. Lett. \textbf{39}, 5892 (2014)}. 

\bibitem{xia-ol16} S. Xia, Y. Hu., D. Song, Y. Zong, L. Tang, and Z. Chen, 
\href{http://dx.doi.org/10.1364/OL.41.001435}
{Opt. Lett. \textbf{41}, 1435 (2016)}. 

\bibitem{zong-oe16} Y. Zong, S. Xia, L. Tang, D. Song, Y. Hu, Y. Pei, J. Su, Y. Li, and Z. Chen, 
\href{https://doi.org/10.1364/OE.24.008877}
{Opt. Express \textbf{24}, 8877 (2016)}. 

\bibitem{mukherjee-prl18} S. Mukherjee, M. Di Liberto, P. \"{O}hberg, R. R. Thomson, and N. Goldman, 
\href{https://doi.org/10.1103/PhysRevLett.121.075502}
{Phys. Rev. Lett. \textbf{121}, 075502 (2018)}. 

\bibitem{ajith-prl18} S. Xia, A. Ramachandran, S. Xia, D. Li, X. Liu, L. Tang, Y. Hu, 
D. Song, J. Xu, D. Leykam, S. Flach, and Z. Chen, 
\href{https://doi.org/10.1103/PhysRevLett.121.263902}
{Phys. Rev. Lett. \textbf{121}, 263902 (2018)}. 

\bibitem{jo-prl12} G.-B. Jo, J. Guzman, C. K. Thomas, P. Hosur, A. Vishwanath, and D. M. Stamper-Kurn, 
\href{https://doi.org/10.1103/PhysRevLett.108.045305}
{Phys. Rev. Lett. \textbf{108}, 045305 (2012)}. 

\bibitem{takahashi-sciadv15} S. Taie, H. Ozawa, T. Ichinose, T. Nishio, S. Nakajima, and Y. Takahashi, 
\href{https://doi.org/10.1126/sciadv.1500854}
{Sci. Adv. \textbf{1}, e1500854 (2015)}. 

\bibitem{takahashi-prl17} H. Ozawa, S. Taie, T. Ichinose, and Y. Takahashi, 
\href{https://doi.org/10.1103/PhysRevLett.118.175301}
{Phys. Rev. Lett. \textbf{118}, 175301 (2017)}. 

\bibitem{masumoto-njp12} N. Masumoto, N. Y. Kim, T. Byrnes, K. Kusudo, A. L\"{o}ffler, 
S. H\"{o}fling, A. Forchel, and Y. Yamamoto, 
\href{http://dx.doi.org/10.1088/1367-2630/14/6/065002}
{New J. Phys. \textbf{14}, 065002 (2012)}. 

\bibitem{baboux-prl16} F. Baboux, L. Ge, T. Jacqmin, M. Biondi, E. Galopin, A. Lema\^{i}tre, 
L. Le Gratiet, I. Sagnes, S. Schmidt, H. E. T\"{u}reci, A. Amo, and J. Bloch, 
\href{https://doi.org/10.1103/PhysRevLett.116.066402}
{Phys. Rev. Lett. \textbf{116}, 066402 (2016)}. 

\bibitem{chen-jmca18} H. Chen, S. Zhang, W. Jiang, C. Zhang, H. Guo, Z. Liu, Z. Wang, F. Liu, and X. Niu,
\href{https://doi.org/10.1039/c8ta02555j}
{J. Mater. Chem. A \textbf{6}, 11252 (2018)}. 

\bibitem{ni-nanoscale18} X. Ni, W. Jiang, H. Huang, K.-H. Jin and F. Liu, 
\href{https://doi.org/10.1039/c8nr02651c}
{Nanoscale \textbf{10}, 11901 (2018)}. 

\bibitem{su-apl18} N. Su, W. Jiang, Z. Wang, and F. Liu, 
\href{https://doi.org/10.1063/1.5017956}
{Appl. Phys. Lett. \textbf{112}, 033301 (2018)}. 

\bibitem{jiang-nanoscale19} W. Jiang, Z. Liu, J.-W. Mei, B. Cui, and F. Liu, 
\href{https://doi.org/10.1039/c8nr08479c}
{Nanoscale \textbf{11}, 955 (2019)}. 

\bibitem{zhang-prb19} S. Zhang, M. Kang, H. Huang, W. Jiang, X. Ni, L. Kang, S. Zhang, H. Xu, Z. Liu, and F. Liu, 
\href{https://doi.org/10.1103/PhysRevB.99.100404}
{Phys. Rev. B \textbf{99}, 100404(R) (2019)}. 

\bibitem{jiang-natcomm19} W. Jiang, H. Huang, and F. Liu, 
\href{https://doi.org/10.1038/s41467-019-10094-3}
{Nat. Commun. \textbf{10}, 2207 (2019)}. 

\bibitem{zhang-aop17} D. Zhang, Y. Zhang, H. Zhong, C. Li, Z. Zhang, Y. Zhang, and M. R. Beli\'{c}, 
\href{https://doi.org/10.1016/j.aop.2017.04.016}
{Ann. Phys. (NY) \textbf{382}, 160 (2017)}. 

\bibitem{franz-prb10} C. Weeks and M. Franz, 
\href{https://doi.org/10.1103/PhysRevB.82.085310}
{Phys. Rev. B \textbf{82}, 085310 (2010)}. 

\bibitem{morais-smith-prb12} W. Beugeling, J. C. Everts, and C. Morais Smith, 
\href{https://doi.org/10.1103/PhysRevB.86.195129}
{Phys. Rev. B \textrm{86}, 195129 (2012)}. 

\bibitem{green-prb10} D. Green, L. Santos, and C. Chamon, 
\href{https://doi.org/10.1103/PhysRevB.82.075104}
{Phys. Rev. B \textbf{82}, 075104 (2010)}. 

\bibitem{nagaosa-prb00} K. Ohgushi, S. Murakami, and N. Nagaosa, 
\href{https://doi.org/10.1103/PhysRevB.62.R6065}
{Phys. Rev. B \textbf{62}, R6065(R) (2000)}. 

\bibitem{haldane-prl04} F. D. M. Haldane, 
\href{https://doi.org/10.1103/PhysRevLett.93.206602}
{Phys. Rev. Lett. \textbf{93}, 206602 (2004)}. 

\bibitem{chen-jpcm12} M. Chen and S. Wan, 
\href{https://doi.org/10.1088/0953-8984/24/32/325502}
{J. Phys.: Condens. Matter \textbf{24}, 325502 (2012)}. 

\bibitem{haldane-prl06} D. N. Sheng, Z. Y. Weng, L. Sheng, and F. D. M. Haldane,
\href{https://doi.org/10.1103/PhysRevLett.97.036808}
{Phys. Rev. Lett. \textbf{97}, 036808 (2006)}. 

\bibitem{zhang-pla16} D.-W. Zhang and S. Cao, 
\href{https://doi.org/10.1016/j.physleta.2016.08.035}
{Phys. Lett. A \textbf{380}, 3541 (2016)}. 

\bibitem{chen-jpa14} L. Chen, T. Mazaheri, A. Seidel, and X. Tang, 
\href{https://doi.org/10.1088/1751-8113/47/15/152001}
{J. Phys. A: Math. Theor. \textbf{47}, 152001 (2014)}. 

\bibitem{cm-smith-prb12} W. Beugeling, N. Goldman, and C. Morais Smith, 
\href{https://doi.org/10.1103/PhysRevB.86.075118}
{Phys. Rev. B \textbf{86}, 075118 (2012)}. 

\bibitem{bloch-rpm08} I. Bloch, J. Dalibard, and W. Zwerger, 
\href{https://doi.org/10.1103/RevModPhys.80.885}
{Rev. Mod. Phys. \textbf{80}, 885 (2008)}. 

\bibitem{liu-pra10} G. Liu, S.-L. Zhu, S. Jiang, F. Sun, and W. M. Liu, 
\href{https://doi.org/10.1103/PhysRevA.82.053605}
{Phys. Rev. A \textbf{82}, 053605 (2010)}. 

\bibitem{shao-prl08} L. B. Shao, S.-L. Zhu, L. Sheng, D. Y. Xing, and Z. D. Wang,
\href{https://doi.org/10.1103/PhysRevLett.101.246810}
{Phys. Rev. Lett. \textbf{101}, 246810 (2008)}. 

\bibitem{hofstetter-pra12} M. Buchhold, D. Cocks, and W. Hofstetter, 
\href{https://doi.org/10.1103/PhysRevA.85.063614}
{Phys. Rev. A \textbf{85}, 063614 (2012)}. 

\bibitem{goldman-prl12} N. Goldman, J. Beugnon, and F. Gerbier, 
\href{https://doi.org/10.1103/PhysRevLett.108.255303}
{Phys. Rev. Lett. \textbf{108}, 255303 (2012)}. 

\bibitem{esslinger-nature14} G. Jotzu, M. Messer, R. Desbuquois, M. Lebrat, T. Uehlinger, D. Greif, and T. Esslinger,
\href{https://doi.org/10.1038/nature13915}
{Nature (London) \textbf{515}, 237 (2014)}. 

\bibitem{goldman-prl10} N. Goldman, I. Satija, P. Nikolic, A. Bermudez, M. A. Martin-Delgado, M. Lewenstein, and I. B. Spielman,
\href{https://doi.org/10.1103/PhysRevLett.105.255302}
{Phys. Rev. Lett. \textbf{105}, 255302 (2010)}. 

\bibitem{monica-prl13} M. Aidelsburger, M. Atala, M. Lohse, J. T. Barreiro, B. Paredes, and I. Bloch,
\href{https://doi.org/10.1103/PhysRevLett.111.185301}
{Phys. Rev. Lett. \textbf{111}, 185301 (2013)}. 

\bibitem{goldman-pra11} N. Goldman, D. F. Urban, and D. Bercioux, 
\href{https://doi.org/10.1103/PhysRevA.83.063601}
{Phys. Rev. A \textbf{83}, 063601 (2011)}. 

\end{thebibliography}
\end{document}